\documentclass[11pt,leqno]{article}

\usepackage{latexsym}
\usepackage{amsfonts}
\usepackage{amsmath}
\usepackage{amssymb}

\makeatletter%
\def\nottoobig#1{{\hbox{$\left#1\vcenter to1.111\ht\strutbox{}\right.\n@space$}}}
\makeatother%

\makeatother%

\makeatletter%

\newcount\hour  \newcount\minutes  \hour=\time  \divide\hour by 60
\minutes=\hour  \multiply\minutes by -60  \advance\minutes by \time
\def\mmmddyyyy{\ifcase\month\or Jan\or Feb\or Mar\or Apr\or May\or Jun\or Jul\or
  Aug\or Sep\or Oct\or Nov\or Dec\fi \space\number\day, \number\year}
\def\hhmm{\ifnum\hour<10 0\fi\number\hour :%
  \ifnum\minutes<10 0\fi\number\minutes}
\def\Draft{{\it Draft of \mmmddyyyy}}

\def\ps@jtsheadings{%
\def\@oddhead{\it\rightmark\hfil\rm\thepage}%
\def\@oddfoot{\hfil\Draft}%
\if@twoside%
\def\@evenhead{\rm\thepage\hfil\it\leftmark}%
\def\@evenfoot{\Draft\hfil}%
\else
\let\@evenhead\@oddhead%
\let\@evenfoot\@oddfoot%
\fi%
}
\def\ps@jtsplain{%
\def\@oddhead{\hfil\Draft}%
\def\@oddfoot{\hfil\rm\thepage\hfil}%
\let\@evenfoot\@oddfoot%
\if@twoside \def\@evenhead{\Draft\hfil} \else \let\@evenhead\@oddhead \fi
}

\def\chaptermark#1{\markboth{\thechapter.\ #1}{\thechapter.\ #1}}%
\def\sectionmark#1{\markright{\thesection.\ #1}}

\def\section{\@startsection {section}{1}{\z@}
    {3.5ex plus1ex minus.2ex}{2.3ex plus.2ex}{\Large\bf}}
\def\subsection{\@startsection{subsection}{2}{\z@}
    {3.25ex plus1ex minus.2ex}{1.5ex plus.2ex}{\large\bf}}
\def\subsubsection{\@startsection{subsubsection}{3}{\z@}
    {3.25ex plus1ex minus.2ex}{1.5ex plus.2ex}{\normalsize\bf}}
\def\paragraph{\@startsection{paragraph}{4}{\z@}
    {3.25ex plus1ex minus.2ex}{1em}{\normalsize\bf}}
\def\subparagraph{\@startsection{subparagraph}{4}{\parindent}
    {3.25ex plus1ex minus.2ex}{1em}{\normalsize\bf}}

\makeatother%

\makeatletter \@beginparpenalty=10000 \makeatother

\def\underl#1 {\leavevmode\let\first=\relax\underli #1 }
\def\underli#1 {\ifx&#1\let\next=\relax\unskip
                \else\let\next=\underli\first\ulinebox{#1}\fi\let\first=\undersp\next}
\def\undersp{\penalty50\ulinebox{\space}\penalty50}
\def\ulinebox#1{\vtop{\hbox{\strut#1}\hrule}}%
\def\unice#1 {\underl #1 & }
\def\desclabel#1{\bf #1\hfil}
\def\desc{\list{}{%
\labelwidth= \leftmargin
\advance \labelwidth by -\labelsep
\let \makelabel=\desclabel}}



\makeatletter %

\newlength{\filength}
\newsavebox{\gcbox}
\sbox{\gcbox}{\framebox[\filength]{\rule{0ex}{2ex}}}

\newlength{\leftjustindent}
\newlength{\@leftjustindent}
\def\leftjust{\let\\\@leftjustcr\let\end\@endleftjust
  \addtolength{\@leftjustindent}{\leftjustindent} \vcenter\bgroup
\halign\bgroup \hbox to\displaywidth{
\rule{\@leftjustindent}{0ex}$\displaystyle##$\hfill }\crcr }
\def\endleftjust{\crcr\egroup\egroup\endgroup}
\def\@endleftjust#1{\crcr\egroup\egroup\@checkend{#1}\endgroup}
\def\@leftjustcr{\crcr}

\newtheorem{theorem}{Theorem}[section]

\newcommand{\qedblob}{\mbox{\rule[-1.5pt]{5pt}{10.5pt}}}
\def\literalqed{{\ \nolinebreak\hfill\mbox{\qedblob\quad}}}

\def\qed{\literalqed}

\newtheorem{proposition}[theorem]{Proposition}
\newcommand{\singlespacing}{\let\CS=
\@currsize\renewcommand{\baselinestretch}{1}\tiny\CS}
\newcommand{\singlespacingplus}{\let\CS=
\@currsize\renewcommand{\baselinestretch}{1.25}\tiny\CS}
\newcommand{\doublespacing}{\let\CS=
\@currsize\renewcommand{\baselinestretch}{1.75}\tiny\CS}
\newcommand{\draftspacing}{\let\CS=
\@currsize\renewcommand{\baselinestretch}{2.0}\tiny\CS}

\makeatother%

\mathcode`\0="0030      %
\mathcode`\1="0031
\mathcode`\2="0032
\mathcode`\3="0033
\mathcode`\4="0034
\mathcode`\5="0035
\mathcode`\6="0036
\mathcode`\7="0037
\mathcode`\8="0038
\mathcode`\9="0039
\singlespacingplus

\newtheorem{definition}[theorem]{Definition}

\makeatletter
\clubpenalty=\@highpenalty
\widowpenalty=\@highpenalty
\makeatother

\makeatletter
\newcommand{\foospacing}{\let\CS=
\@currsize\renewcommand{\baselinestretch}{1.15}\tiny\CS}
\newcommand{\niceonespacing}{\let\CS=\@currsize\renewcommand{\baselinestretch}{1.1}\tiny\CS}\newcommand{\nicetwospacing}{\let\CS=\@currsize\renewcommand{\baselinestretch}{1.2}\tiny\CS}
\newcommand{\nicethreespacing}{\let\CS=\@currsize\renewcommand{\baselinestretch}{1.3}\tiny\CS}
\newcommand{\singlespacingplusplus}{\let\CS=\@currsize\renewcommand{\baselinestretch}{1.35}\tiny\CS}
\newcommand{\nicefourspacing}{\let\CS=\@currsize\renewcommand{\baselinestretch}{1.4}\tiny\CS}
\newcommand{\nicefivespacing}{\let\CS=\@currsize\renewcommand{\baselinestretch}{1.5}\tiny\CS}
\newcommand{\nicesixspacing}{\let\CS=\@currsize\renewcommand{\baselinestretch}{1.6}\tiny\CS}
\makeatother

\makeatletter
\def\@cite#1#2{[#1\if@tempswa , #2\fi]}
\makeatother

\makeatletter
\def\@citex[#1]#2{\if@filesw\immediate\write\@auxout{\string\citation{#2}}\fi
  \def\@citea{}\@cite{\@for\@citeb:=#2\do
    {\@citea\def\@citea{,\linebreak[0]}\@ifundefined
       {b@\@citeb}{{\bf ?}\@warning
       {Citation `\@citeb' on page \thepage \space undefined}}%
\hbox{\csname b@\@citeb\endcsname}}}{#1}}
\makeatother

\makeatletter
\def\ps@thesis{\def\@oddhead{\hfil\rm\thepage\hfil}\def\@oddfoot{}\def\@evenhead{\hfil\rm\thepage\hfil}\def\@evenfoot{}\def\chaptermark##1{}\def\sectionmark##1{}}
\makeatother

\makeatletter
\def\foobarpt{\textfont\z@\tenrm 
  \scriptfont\z@\ninrm \scriptscriptfont\z@\sevrm
\textfont\@ne\tenmi \scriptfont\@ne\ninmi \scriptscriptfont\@ne\sevmi
\textfont\tw@\tensy \scriptfont\tw@\ninsy \scriptscriptfont\tw@\sevsy
\textfont\thr@@\tenex \scriptfont\thr@@\tenex \scriptscriptfont\thr@@\tenex
\def\unboldmath{\everymath{}\everydisplay{}\@nomath\unboldmath
          \textfont\@ne\tenmi 
          \textfont\tw@\tensy \textfont\lyfam\tenly
          \@boldfalse}\@boldfalse
\def\boldmath{\@ifundefined{tenmib}{\global\font\tenmib\@mbi\@magscale1\global
        \font\tensyb\@mbsy \@magscale1\global\font
         \tenlyb\@lasyb\@magscale1\relax\@addfontinfo\@xiipt
              {\def\boldmath{\everymath
                {\mit}\everydisplay{\mit}\@prtct\@nomathbold
                \textfont\@ne\tenmib \textfont\tw@\tensyb 
                \textfont\lyfam\tenlyb\@prtct\@boldtrue}}}{}\@xiipt\boldmath}%
\def\prm{\fam\z@\tenrm}%
\def\pit{\fam\itfam\tenit}\textfont\itfam\tenit \scriptfont\itfam\ninit
   \scriptscriptfont\itfam\sevit
\def\psl{\fam\slfam\tensl}\textfont\slfam\tensl 
     \scriptfont\slfam\tensl \scriptscriptfont\slfam\tensl
\def\pbf{\fam\bffam\tenbf}\textfont\bffam\tenbf 
   \scriptfont\bffam\ninbf \scriptscriptfont\bffam\ninbf 
\def\ptt{\fam\ttfam\tentt}\textfont\ttfam\tentt
   \scriptfont\ttfam\nintt \scriptscriptfont\ttfam\nintt 
\def\psf{\fam\sffam\tensf}\textfont\sffam\tensf
    \scriptfont\sffam\tensf \scriptscriptfont\sffam\tensf
\def\psc{\@getfont\psc\scfam\@xiipt{\@mcsc\@magscale1}}%
\def\ly{\fam\lyfam\tenly}\textfont\lyfam\tenly 
   \scriptfont\lyfam\ninly \scriptscriptfont\lyfam\sevly
 \@setstrut \rm}

\makeatother

\newcommand{\up}{{\rm UP}}

\newcommand{\p}{{\rm P}}

\newcommand{\np}{{\rm NP}}

\newcommand{\pair}[1]{\mathopen\langle{#1}\mathclose\rangle}

\newcommand{\sigmastar}{\ensuremath{\Sigma^\ast}}

\newcommand{\pisnotnp}{\ensuremath{\p\neq\np}}

\newcommand{\condition}{\,\nottoobig{|}\:}

\newcommand{\seq}{\subseteq}

\newcommand{\fp}{{\rm FP}}

\def\equalsdef{=}
\def\equalsdeffootnote{=}
\newcommand{\weakaowf}{A$^{\!\!\mbox{\protect\scriptsize w}\!}$OWF}

\newcommand{\weakaowfs}{A$^{\!\!\mbox{\protect\scriptsize w}\!}$OWFs}
\newcommand{\weakaowfsmath}{A$^{\!\mbox{\protect\scriptsize w}\!}$OWFs}
\newcommand{\weakaowffootnote}{A$^{\!\!\mbox{\protect\tiny w}\!}$OWF}
\newcommand{\weakaowfsfootnote}{A$^{\!\!\mbox{\protect\tiny w}\!}$OWFs}

\newenvironment{block}{\begin{list}{\hbox{}}{\leftmargin 1em
    \itemindent -1em \topsep 0pt \itemsep 0pt \partopsep 0pt}}{\end{list}}

\title{
Creating Strong Total Commutative Associative
Complexity-Theoretic
One-Way Functions from Any
Complexity-Theoretic One-Way Function
}

\author{
Lane A. Hemaspaandra\,\thanks{
Email: {\tt lane@cs.rochester.edu}.
Supported in part 
by grants NSF-CCR-9322513 and 
NSF-INT-9513368/DAAD-315-PRO-fo-ab.}
\\
Department of Computer Science\\
University of Rochester\\
Rochester, NY 14627, USA\\
\and 
J\"{o}rg Rothe\,\thanks{
Email: {\tt rothe@informatik.uni-jena.de}.
Supported in part 
by grant
NSF-INT-9513368/DAAD-315-PRO-fo-ab and a 
NATO Postdoctoral Science Fellowship
from the Deut\-scher Aka\-de\-mi\-scher Aus\-tausch\-dienst
(``Ge\-mein\-sames Hoch\-schul\-sonder\-pro\-gramm~III 
von Bund und L\"andern'').
Work done in part while visiting the 
University of Rochester.}
\\
Institut f\"ur Informatik\\
Friedrich-Schiller-Universit\"at Jena\\
07740 Jena, Germany}

\date{}
\makeatletter
\def\@listI{\leftmargin\leftmargini \parsep 4.5pt plus 1pt minus 1pt\topsep
6pt plus 2pt minus 2pt \itemsep  2pt plus 2pt minus 1pt}

\let\@listi\@listI
\@listi
\makeatother

\begin{document}

\typeout{WARNING:  BADNESS used to suppress reporting!  Beware!!}
\hbadness=3000%
\vbadness=10000 %

\bibliographystyle{alpha}

\pagestyle{empty}
\setcounter{page}{1}

\sloppy

\pagestyle{empty}
\setcounter{footnote}{0}

{\singlespacing

\maketitle

}

\begin{center}
{\large\bf Abstract}
\end{center}
\begin{quotation}
{\singlespacing 
  
Rabi and Sherman~\cite{rab-she:j:aowf} presented novel digital
signature and unauthenticated secret-key agreement protocols,
developed by
themselves and by Rivest and Sherman.  These 
protocols use ``strong,'' total,
commutative (in the case of multi-party secret-key agreement),
associative one-way functions as their key building blocks.  Though
Rabi and Sherman did prove that
associative one-way
functions exist
if $\p \neq \np$, they left as an open question whether any natural
complexity-theoretic assumption is sufficient to ensure the
existence of ``strong,'' total, commutative, 
associative one-way functions.  In
this paper, we prove that if $\p \neq \np$ then ``strong,'' total,
commutative, associative one-way functions exist. 

}

\vspace*{2mm}
{\singlespacing 

}
\end{quotation}

\foospacing
\setcounter{page}{1}
\pagestyle{plain}
\sloppy

\section{Introduction and Preliminaries}

Rabi and Sherman~\cite{rab-she:j:aowf}
study associative one-way functions (AOWFs) 
and show that AOWFs exist exactly if $\p \neq \np$.
They also present the notion of strong AOWFs---AOWFs 
that are hard to invert even when one of their
arguments is given. 
They give protocols due to 
Rivest and Sherman
for two-party secret-key agreement and due to 
Rabi and Sherman for digital signatures,
that depend on
strong, total AOWFs.
They also outline
a protocol approach
for multi-party secret-key agreement
that depends on strong, total, commutative AOWFs.  

There are two key worries regarding the Rabi-Sherman approach.  The
first is whether their protocols are secure even if strong, total,
commutative AOWFs exist.  This worry has two facets.  The first facet
is that, as they note, 
like 
Diffie-Hellman~\cite{dif-hel:j:diffie-hellman,dif-hel:j:intro-privacy} 
the protocol they describe
has no current proof of security (even if the existence of strong,
total, commutative AOWFs is given), though
Rabi and Sherman give intuitively
attractive arguments suggesting the plausibility of security.  
In particular, they prove that certain direct attacks against 
their protocols are precluded by the fact that the protocols
use strong, total AOWFs as building blocks.
The
second facet of the first worry is that their definition of strong,
total, commutative AOWFs is a worst-case definition, as opposed to the
average-case definition one desires for a satisfyingly strong
approach to cryptography.  

The second worry is that Rabi and Sherman provide no evidence at
all that strong, total, commutative AOWFs exist, though they do prove
that AOWFs exist if $\pisnotnp$.\footnote{\protect\singlespacing 
We mention that, after
we sent this paper to them, they (Sherman, 
personal communication, June 1998)
informed us
that they 
had had discussions and proof sketches towards achieving
the claim that 
strong AOWFs exist
if $\p\neq\np$.}
In this paper we completely remove
that worry by proving that strong, total, commutative AOWFs exist if
$\pisnotnp$.  (In light of the above-mentioned first
worry---and especially its second facet---we note, as did Rabi and
Sherman, that the study of AOWFs should be viewed as more of
complexity-theoretic interest than of applied cryptographic
interest, though it is hoped that AOWFs will in the long term
prove, probably in average-case versions, to be of 
substantial applied cryptographic
value.)
    
Phrasing our work in a slightly different but 
equivalent way, in this paper we prove that the existence of AOWFs 
(or, indeed, the existence of 
{\em any\/} one-way function)
implies
the existence of strong, total, commutative  AOWFs.  
Furthermore, based on Kleene's~\cite{kle:b:metamathematics}
distinction between {\em weak\/} 
and {\em complete equality\/} of partial 
functions, 
we give a definition of associativity that, for partial functions, is
a more natural analog of the standard total-function definition
than that of Rabi and Sherman, and
we show that 
their and our results hold even under this more natural definition.

Fix the alphabet $\Sigma = \{0,1\}$, and let $\sigmastar$ denote the
set of all strings over~$\Sigma$.  The length of any string $x \in
\sigmastar$ will be denoted by~$|x|$. Throughout this paper, when we
use ``binary function'' we mean ``two-argument function.'' Unless
explicitly stated as being total, all functions may potentially be partial,
i.e., ``let $\sigma$ be any binary function'' does not imply that
$\sigma$ will necessarily be total.
For any binary function
$\sigma$,
we will interchangeably use prefix and infix notation,
i.e., $\sigma(x,y) = x \sigma y$. 
As is standard, pairs of strings will sometimes be encoded as a single
string by some standard total, one-to-one, onto, polynomial-time
computable pairing function, $\pair{\cdot, \cdot} : \sigmastar \times
\sigmastar \rightarrow\, \sigmastar$, that has polynomial-time
computable inverses, and is non-decreasing in each argument when the
other argument is fixed.  Let FP denote the set of all polynomial-time
computable (partial) functions.  Regarding Part~\ref{p:one-way} of the
following definition, we mention that we use the
term ``one-way function'' in the same way Rabi and
Sherman~\cite{rab-she:j:aowf} do, i.e., in the complexity-theoretic
(that is, worst-case) sense, and without 
requiring that the function
necessarily be injective.

\begin{definition} 
\label{def:one-way}
Let $\sigma : \sigmastar \times \sigmastar \rightarrow\, \sigmastar$ 
be any binary function.
{\singlespacing
\begin{enumerate}
\item We say $\sigma$ is {\em honest\/} if and only if there exists some
  polynomial $p$ such that for every $z \in \mbox{\rm range}(\sigma)$
  there exists a pair $(x,y) \in \mbox{\rm domain}(\sigma)$ such that $x
  \sigma y = z$ and $|x| + |y| \leq p(|z|)$.\footnote{\protect\singlespacing 
This definition of honesty for binary functions is that of
Rabi and Sherman~\cite{rab-she:j:aowf}, 
and is equivalent to requiring $|\pair{x,y}| \leq p(|z|)$, since there
exists some polynomial $q$ (that depends on the pairing function chosen) 
such that for every $x,y \in \sigmastar$,
$|\pair{x,y}| \leq q(|x| + |y|)$ and $|x| + |y|\leq q(|\pair{x,y}|)$.
}

\item We say $\sigma$ is {\em $\fp$-invertible\/} if and only if there
  exists a total function $g \in \fp$ such that for every $z \in \mbox{\rm
  range}(\sigma)$, $g(z)$ is some element of
  $\sigma^{-1}(z) \equalsdef \{ (x,y) \in \mbox{\rm domain}(\sigma)
  \condition x \sigma y = z \}$.

\item \label{p:one-way}
We say $\sigma$ is a {\em one-way function\/} if and only if
  $\sigma$ is honest, polynomial-time computable, and not
  $\fp$-invertible.
\end{enumerate}
}
\end{definition}

Rabi and Sherman~\cite{rab-she:j:aowf} 
define a notion of associativity for binary functions as follows:

\begin{definition} 
\label{def:rs-associative}
Let $\circ : \sigmastar \times
\sigmastar \rightarrow\, \sigmastar$ be any binary
function. We say $\circ$ is 
{\em weakly associative\/}\footnote{\protect\singlespacing
They call this ``associative,'' but for reasons we will immediately 
make clear, we use ``weakly associative'' to describe their notion.
}
if and only if $x \circ (y \circ z) = (x \circ y) \circ z$ holds
for all $x,y,z \in \sigmastar$ such that 
each of $(x,y)$, $(y,z)$, $(x,y \circ z)$,
and $(x \circ y, z)$ is an element of $\mbox{\rm domain}(\circ)$.
\end{definition} 

This type of associativity, however, is not natural 
for non-total functions, since it does not evaluate as being 
false ``equations'' such as 
``$\mbox{\rm undefined} = 1010$'' 
(this can occur in $x \circ (y \circ z) = (x \circ y) \circ z$ 
in various ways, e.g., if 
$(x,y)$, $(x \circ y, z)$, and $(y,z)$ are in the domain of $\circ$
but $(x,y \circ z)$ is not).
It would seem more natural for a definition of associativity 
for binary functions to
require that both sides of the above equation stand or fall together.
That is, for each triple of strings $x,y,z \in \sigmastar$, 
either both sides should be 
defined and equal, or each side should be undefined. Drawing on Kleene's
careful discussion 
of how to define equality between partial functions, our 
definition of associativity---given in Definition~\ref{def:associative}
below---achieves this natural behavior.

Associativity expresses equality between two functions each of
which can be viewed as a 3-ary function that results from a
given binary function. The distinction in the two definitions
of associativity can be said to come from two distinct interpretations
of ``equality'' between functions, known in
recursive function theory as {\em weak equality\/} and 
{\em complete equality}
(see Kleene~\cite{kle:b:metamathematics}). Kleene suggests the use of two
different equality symbols---we will use ``$=_w$'' and ``$=_c$'' and we 
have modified the following quotation to use these also---and he writes:
\begin{quote}
\noindent
We now introduce ``$\psi(x_1, \ldots , x_n) =_c \chi(x_1, \ldots ,
x_n)$'' to express, for particular $x_1, \ldots , x_n$, that if either
of $\psi(x_1, \ldots , x_n)$ and $\chi(x_1, \ldots , x_n)$ is defined,
so is the other and the values are the same (and hence if either of
$\psi(x_1, \ldots , x_n)$ and $\chi(x_1, \ldots , x_n)$ is undefined,
so is the other).  The difference in the meaning of (i)~``$\psi(x_1,
\ldots , x_n) =_w \chi(x_1, \ldots , x_n)$'' and (ii)~``$\psi(x_1,
\ldots , x_n) =_c \chi(x_1, \ldots , x_n)$'' comes when one of
$\psi(x_1, \ldots , x_n)$ and $\chi(x_1, \ldots , x_n)$ is
undefined. Then (i) is undefined, while (ii) is true or false
according as the other is or is not
undefined.---~\cite[pp.~327--328]{kle:b:metamathematics}
\end{quote}  
\noindent
We feel that
complete equality is the more natural of the two notions.
Thus, following the notion of {\em complete
equality\/} between functions, 
we propose the following definition
of associativity for binary functions. Nonetheless, we will
show that 
the results of Rabi and Sherman~\cite{rab-she:j:aowf} 
and of the present paper
hold even under this more restrictive definition.  In a similar vein,
we also define commutativity for (partial) binary functions.

\begin{definition} 
\label{def:associative}
Let $\sigma : \sigmastar \times \sigmastar \rightarrow\, \sigmastar$
be any binary function.  Define $\Gamma \equalsdef \sigmastar
\cup \{\bot\}$ 
and define an extension $\widehat{\sigma} : \Gamma \times
\Gamma \rightarrow\, \Gamma$
of $\sigma$ as follows:
\[
  \widehat{\sigma}(a,b) \equalsdef \left\{ 
\begin{array}{ll} \sigma(a,b)  &
  \mbox{if $a \neq \bot$ and $b \neq \bot$ and $(a,b) \in \mbox{\rm
  domain}(\sigma)$} \\ 
  \bot & \mbox{otherwise.}
\end{array} 
\right.
\]
We say $\sigma$ is {\em associative\/} if and only if, for every
$x,y,z \in \sigmastar$, $(x \widehat{\sigma} y) \widehat{\sigma} z = x
\widehat{\sigma} (y \widehat{\sigma} z)$.
We say $\sigma$ is {\em commutative\/} if and only if, for every
$x,y \in \sigmastar$, $x \widehat{\sigma} y = y \widehat{\sigma} x$ 
$($i.e., $x \sigma y  =_c y \sigma x )$.
\end{definition}

Clearly, every associative function is weakly associative,
since our notion of associativity is more restrictive
than weak associativity. The converse, however, is not always true,
so these are indeed different notions.

\begin{proposition}
\label{prop:ass}
{\singlespacing
\begin{enumerate}
\item \label{prop:ass1} Every associative binary function is
weakly associative.

\item \label{prop:ass2} Every total binary function is associative if
and only if it is weakly associative.

\item \label{prop:ass3} There exists a binary function that is
weakly associative, but not associative.
\end{enumerate}
}
\end{proposition}

\noindent
{\bf Proof.} (\ref{prop:ass1}) and (\ref{prop:ass2}) are immediate
from the definitions.
To prove (\ref{prop:ass3}), we define the following binary
function $\sigma : \sigmastar \times \sigmastar \rightarrow\,
\sigmastar$:
\[
\sigma(a,b) \equalsdef \left\{ 
\begin{array}{ll} 
  111  & \mbox{if $a = 1$ and $b = 11$} \\ 

  0    & \mbox{if $a = 111$ and $b = 1111$} \\

  \mbox{\rm undefined} & \mbox{otherwise.}
\end{array} 
\right.
\]
By ``undefined'' above we do not mean some new token ``undefined,'' 
but rather we simply mean that for cases handled by that line of the 
definition $(a,b) \not\in \mbox{\rm domain}(\sigma)$.

Let $\widehat{\sigma}$ be the extension of $\sigma$ defined in
Definition~\ref{def:associative}. Note that $(1 \widehat{\sigma} 11)
\widehat{\sigma} 1111 = 0$, but $1 \widehat{\sigma} (11
\widehat{\sigma} 1111) = \bot$.
Thus, $\sigma$ is not associative. However, $\sigma$
{\em is\/} weakly associative, since no three strings in $\sigmastar$
satisfy the four domain conditions required in
Definition~\ref{def:rs-associative}.~\qed

\begin{definition} 
{\singlespacing
\begin{enumerate}
\item A binary function $\sigma : \sigmastar \times \sigmastar
\rightarrow\, \sigmastar$ is an {\em AOWF\/} if and only if $\sigma$
is both associative and a one-way function.

\item {\rm \cite{rab-she:j:aowf}}\quad 
A binary function $\sigma : \sigmastar \times \sigmastar
\rightarrow\, \sigmastar$ is an {\em \weakaowf\/} if and only if 
$\sigma$ is both weakly associative and a one-way function.
\end{enumerate}
}
\end{definition} 

Rabi and Sherman~\cite{rab-she:j:aowf} also introduce the notion of
{\em strong\/} one-way functions---binary one-way functions that are hard 
to invert even if one of their arguments is given. 
Strongness clearly implies one-way-ness.
(We note that
``strongness'' here should not be confused with the property of 
strong-FP-invertibility of functions 
introduced by
Allender~\cite{all:coutdatedExceptForPUNCstuff:complexity-sparse,all:thesis:invertible}.)
To avoid any possibility of ambiguity we henceforward, when using
equality signs with partial functions, will make it explicit that by
equality we mean~$=_c$.

\begin{definition} 
\label{def:strong}
A binary function
$\sigma$ is said to be {\em strong\/} 
if and only if 
$\sigma$ is not $\fp$-invertible even if
one of its arguments is given. 
More formally, binary function $\sigma$ is {\em strong\/} 
if and only if neither (a) nor (b) holds:
\begin{description}
\item[(a)] There exists a total function $g_1 \in \fp$ such that for every
  $z \in \mbox{\rm range}(\sigma)$ and for each $x \in \sigmastar$, if
  $\sigma(x,y) =_c z$ for some $y \in \sigmastar$, then $\sigma(x,
  g_1(\pair{x,z})) =_c z$.
\item[(b)] There exists a total function $g_2 \in \fp$ such that for every
  $z \in \mbox{\rm range}(\sigma)$ and for each $y \in \sigmastar$, if
  $\sigma(x,y) =_c z$ for some $x \in \sigmastar$, then 
  $\sigma(g_2(\pair{y,z}), y) =_c z$.
\end{description}
\end{definition}

\section{Main Result}

Rabi and Sherman~\cite{rab-she:j:aowf} show that 
\weakaowfs\
exist if and only if $\p \neq \np$. 
They 
present no evidence that {\em strong\/}
\weakaowfs\
exist, and they establish no structural conditions sufficient to imply
that any exist.
Solving these open questions, we show in Theorem~\ref{thm:aowf-equ} below 
that there exist strong, total, commutative 
\weakaowfs\ 
(equivalently, strong, total, commutative 
AOWFs) if and only if $\p \neq \np$.

\begin{theorem}
\label{thm:aowf-equ}
The following are equivalent.
{\singlespacing
\begin{enumerate}
\item \label{thm:aowf-equ1} $\p \neq \np$.
\item \label{thm:aowf-equ2} There exist \weakaowfsmath.
\item \label{thm:aowf-equ3} There exist AOWFs.
\item \label{thm:aowf-equ4} There exist strong, total, commutative \weakaowfsmath.
\item \label{thm:aowf-equ5} There exist strong, total, commutative AOWFs.
\end{enumerate}
}
\end{theorem}

\noindent
{\bf Proof.} By Proposition~\ref{prop:ass}.\ref{prop:ass2}, 
(\ref{thm:aowf-equ4}) and (\ref{thm:aowf-equ5}) are equivalent.
Rabi and Sherman~\cite{rab-she:j:aowf} have shown the
equivalence of (\ref{thm:aowf-equ1}) and (\ref{thm:aowf-equ2}), by
exploiting the associativity of the closest common ancestor relation
for configurations in the computation tree of nondeterministic Turing
machines. 
Since (\ref{thm:aowf-equ5}) (and, equivalently, (\ref{thm:aowf-equ4})) 
implies (\ref{thm:aowf-equ2}) and (\ref{thm:aowf-equ3}), and since each of 
(\ref{thm:aowf-equ2}) and (\ref{thm:aowf-equ3}) implies
(\ref{thm:aowf-equ1}) (by Proposition~\ref{prop:ass}.\ref{prop:ass1}
and by the equivalence of (\ref{thm:aowf-equ1}) and (\ref{thm:aowf-equ2})),
it suffices to show that (\ref{thm:aowf-equ1}) implies 
(\ref{thm:aowf-equ5}) to establish the theorem.

Assume $\p \neq \np$, and let $A$ be a set in $\np - \p$. Let $M$ be
a nondeterministic polynomial-time Turing machine accepting~$A$. By a
{\em witness\/} for ``$x \in A$'' we mean a string $w \in
\sigmastar$ that encodes some
accepting computation path of $M$ on input~$x$.  Assume, without loss
of generality,
that for each $x \in A$, every witness $w$ certifying that $x \in A$
satisfies $|w| = p(|x|) > |x|$ for some strictly increasing polynomial
$p$ depending on~$M$. For each string~$x$, define the set of
witnesses for ``$x \in A$'' (with respect to~$M$) by
\[
W_M(x) \equalsdef \{ w \condition w \mbox{ is a witness for ``$x
\in A$''} \} .
\]
Note that if $x \not\in A$ then 
$W_M(x) = \emptyset$.

For any strings $u,v,w\in\sigmastar$, 
$\min(u,v)$ will denote 
the lexicographically smaller of~$u$ and~$v$, and 
$\min(u,v,w)$ will denote 
the lexicographically smallest of~$u$, $v$, and~$w$.
Define the binary function $\sigma : \sigmastar \times
\sigmastar \rightarrow\, \sigmastar$ by
\[
\sigma(a,b) \equalsdef \left\{ 
\begin{array}{ll} 
  \pair{x,\min(w,y)} & \mbox{if $(\exists x \in \sigmastar)\, 
                         (\exists w,y \in W_M(x))\, 
                         [a = \pair{x,w} \ \wedge\ b = \pair{x,y}]$} \\ 

  \pair{x,x} & \mbox{if $(\exists x \in \sigmastar)\, 
                         (\exists w \in W_M(x))\, 
                         [(a = \pair{x,x} \ \wedge\ b = \pair{x,w})$} \\ 
             & \hfill 
               \mbox{$\vee\ (a = \pair{x,w} \ \wedge\ b = \pair{x,x})]$} \\

  \mbox{\rm undefined} & \mbox{otherwise.}
\end{array} 
\right.
\]

On our way towards a proof that (\ref{thm:aowf-equ1}) 
implies~(\ref{thm:aowf-equ5}), we will first prove that 
the function $\sigma$ defined above is a strong, commutative
AOWF\@. Then we will show how to extend $\sigma$ to a
strong, total, commutative AOWF, 
thus establishing~(\ref{thm:aowf-equ5}).

$\sigma$ is clearly honest.
Also, $\sigma \in \fp$. 
That is, given $(a,b)$ as the input, it is easy to
decide in polynomial time whether $(a,b) \in \mbox{\rm 
domain}(\sigma)$,
and if so, which of $\pair{x,x}$ or $\pair{x,w}$ for
suitable $x \in \sigmastar$ and $w \in W_M(x)$ should be output as
the value of $\sigma(a,b)$.\footnote{\protect\singlespacing
Recall our assumption that for each~$x \in A$, every witness $w$
for ``$x \in A$'' satisfies $|w| = p(|x|) > |x|$.
This assumption ensures that there is no ambiguity in determining whether
$a$ and $b$ are of the form $\pair{x,x}$ or of the form 
$\pair{x,\mbox{\protect\footnotesize\rm PotentialWitness}}$,
and checking items of the form 
$\pair{x,\mbox{\protect\footnotesize\rm PotentialWitness}}$ is easy as 
$\bigcup_{x \in \sigmastar} W_M(x)$ is in~$\p$.
}

Now, we show that $\sigma$ cannot be inverted in polynomial time,
even if one of its arguments is given. Assume, for instance, that 
there exists a total function $g_2 \in \fp$ such that given any $z$ in the 
range of $\sigma$ and any second argument~$b$ for which there is some
$a \in \sigmastar$ with $\sigma(a,b) =_c z$, it holds that
$\sigma(g_2(\pair{b,z}), b) =_c z$.
Then, contradicting our assumption that $A \not\in \p$, $A$
could be decided in polynomial time as follows. On input~$x$, 
to decide whether or not $x \in A$, compute
$g_2 (\pair{\pair{x,x},\pair{x,x}})$, interpret it as a pair
$\pair{d,e}$, and accept if and only if $d = x$ and $e \in W_M(x)$. 
An analogous proof works for the case of a
fixed first argument. Thus, neither (a) nor (b)
of Definition~\ref{def:strong} holds, so
$\sigma$ is a strong one-way function.

We now prove that $\sigma$ is associative. 
Let $\widehat{\sigma}$ be the extension of $\sigma$ from
Definition~\ref{def:associative}. 
Fix any strings
$a = \pair{a_1, a_2}$, $b = \pair{b_1, b_2}$, and
$c = \pair{c_1, c_2}$ in~$\sigmastar$.
Let $k$ equal how many of 
$a_2$, $b_2$, and~$c_2$
are in $W_M(a_1)$. For example, if $a_2 = b_2 = c_2 \in W_M(a_1)$,
then $k = 3$.
To show that 
\begin{eqnarray}
\label{equ:ass}
(a \widehat{\sigma} b) \widehat{\sigma} c = a \widehat{\sigma} (b
\widehat{\sigma} c)
\end{eqnarray}
holds, we distinguish the following cases.

\begin{description}
\item[Case~1:] $[ a_1 \neq b_1 \ \vee\ a_1 \neq c_1 \ \vee\ b_1 \neq c_1 ]$.
In light of the definition of~$\sigma$,
we have
\begin{eqnarray}
\label{equ:ass-alpha1}
(a \widehat{\sigma} b) \widehat{\sigma} c = \bot = a \widehat{\sigma} (b
\widehat{\sigma} c) .
\end{eqnarray}

\item[Case~2:] $[a_1 = b_1 = c_1 \ \wedge\ 
\{a_2, b_2, c_2 \} \not\seq \{a_1\} \cup W_M(a_1)]$.
(\ref{equ:ass-alpha1}) holds here too, in light 
of the definition of~$\sigma$.

\item[Case~3:] $[a_1 = b_1 = c_1 \ \wedge\ 
\{a_2, b_2, c_2 \} \seq \{a_1\} \cup W_M(a_1)]$.
In this case, note that $\widehat{\sigma}$
decreases by one the number of witnesses, in particular preserving
the lexicographic minimum if both 
arguments contain witnesses for ``$a_1\in A$,'' 
outputting $\pair{a_1, a_1}$ if exactly
one of its arguments contains a witness for ``$a_1 \in A$,''
and outputting $\bot$ if neither contains a witness for ``$a_1 \in A$.''
So it is not hard to see that (in the current case) 
if $k \in \{0,1\}$ then (\ref{equ:ass-alpha1}) holds,
if $k=2$ then
$$
(a \widehat{\sigma} b) \widehat{\sigma} c = \pair{a_1, a_1} = 
a \widehat{\sigma} (b \widehat{\sigma} c) 
$$
holds, and if $k=3$ then
$$
(a \widehat{\sigma} b) \widehat{\sigma} c = \pair{a_1, \min(a_2, b_2, c_2)} = 
a \widehat{\sigma} (b \widehat{\sigma} c) 
$$
holds.
\end{description}

Note that in each case (\ref{equ:ass}) is satisfied.
Furthermore, it is easy to see from the definition of
$\sigma$ that $\sigma$ is commutative.
Thus, $\sigma$ is a strong, commutative AOWF, as claimed earlier. 

Finally, to complete the proof, we now show how to extend
$\sigma$ to a strong, {\em total}, 
commutative
AOWF\@.\footnote{\protect\singlespacing
Rabi and Sherman~\cite{rab-she:j:aowf} give a construction that they
claim lifts any \weakaowffootnote\ whose domain is in P to a total
\weakaowffootnote. However, it is far from clear that their
construction achieves this.  In fact, we show that any proof that
their construction is valid would immediately prove
that $\up = \np$.
(Note:  Valiant's class UP consists of those
languages accepted by nondeterministic polynomial-time Turing machines
having the property that on all inputs they have no more than one
accepting path~\protect\cite{val:j:checking}.)
In particular, we provide the following counterexample
to Rabi and Sherman's assertion,
the proof of which shows that if $\up\neq\np$ 
then their construction does not
always preserve weak associativity.
\begin{proposition}
If $\up\neq\np$, then there exists an {\rm \weakaowffootnote}~$\tilde{\sigma}$,
satisfying $(\exists \tilde{a}) [(\tilde{a},\tilde{a})
\not\in \mbox{\rm domain}(\tilde{\sigma})]$ and having domain in~$\p$, 
such that 
the construction that Rabi and Sherman claim converts 
{\rm \weakaowfsfootnote}~into total {\rm \weakaowfsfootnote}~in fact fails 
on~$\tilde{\sigma}$.
\end{proposition}
We prove the proposition as follows.
Fix a set $A' \in \np - \up$ and an NP machine $M'$ accepting~$A'$.
Let the polynomial $p'$ and, for each~$x$, let the 
witness sets $W_{M'}(x)$ be defined analogous to the definitions of
$p$ and $W_M(x)$ earlier in the proof of Theorem~\ref{thm:aowf-equ}.
Define the binary function $\tilde{\sigma} : \sigmastar \times
\sigmastar \rightarrow\, \sigmastar$ by
\[
\tilde{\sigma}(a,b) \equalsdeffootnote \left\{ 
\begin{array}{ll} 
  \pair{x,w} & \mbox{if $(\exists x \in \sigmastar)\, 
                         (\exists w \in W_{M'}(x))\, 
                         [a = \pair{x,w} = b]$} \\ 

  \pair{x,x} & \mbox{if $(\exists x \in \sigmastar)\, 
                         (\exists w \in W_{M'}(x))\, 
                         [(a = \pair{x,x} \ \wedge\ b = \pair{x,w})$} \\ 
             & \hfill 
               \mbox{$\vee\ (a = \pair{x,w} \ \wedge\ b = \pair{x,x})]$} \\

  \mbox{\rm undefined} & \mbox{otherwise.}
\end{array} 
\right.
\]
It is not hard to verify that $\tilde{\sigma}$ is indeed an
\weakaowffootnote.
Let $\tilde{a}$ be a fixed string such that $(\tilde{a}, \tilde{a}) \not\in
\mbox{\rm domain}(\tilde{\sigma})$.
For the particular function $\tilde{\sigma}$ defined above,
such a string $\tilde{a}$ indeed exists 
(e.g.,~let $\tilde{a} \equalsdeffootnote \pair{x_0,1x_0}$ for any particular
fixed $x_0 \not\in A',$ see the discussion 
of $a_0$ in the proof of Theorem~\ref{thm:aowf-equ}
as to why this is right)---in contrast,
the ``$c$'' of~\cite[p.~242,~l.~10]{rab-she:j:aowf} may not in 
general exist. 
Now, using the Rabi-Sherman technique, 
extend $\tilde{\sigma}$ to a total
function, $\tilde{\tau}$, the same way we will obtain 
the total extension $\tau$ of $\sigma$
later in the proof of Theorem~\ref{thm:aowf-equ}. 
Fix some string $\tilde{x} \in A'$ that has two
distinct witnesses $w$ and $y$ in $W_{M'}(\tilde{x})$
(such $\tilde{x}$, $w$, and $y$ exist, as $A' \not\in \up$), and let 
$a = \pair{\tilde{x},w}$,
$b = \pair{\tilde{x},y}$, and 
$c = \pair{\tilde{x},\tilde{x}}$.  Then, we have
$(a \tilde{\tau} b) \tilde{\tau} c = 
\tilde{a} \neq \pair{\tilde{x},\tilde{x}} = 
a \tilde{\tau} (b \tilde{\tau} c)$,
and thus $\tilde{\tau}$ 
is not associative (and thus, as it is total, is not weakly associative).
(The 
reason that $(a \tilde{\tau} b) \tilde{\tau} c = \tilde{a}$ 
may not be clear to the reader;  to see why this holds, one must 
look at the Rabi-Sherman technique of extending $\tilde{\sigma}$ to 
$\tilde{\tau}$, which, very informally, is to use
$\tilde{a}$ as a dumping ground.)
We mention that, for essentially the same reason, 
$\tilde{\sigma}$ is not associative (and thus is not an 
AOWF), since
$(a \widehat{\tilde{\sigma}} b) \widehat{\tilde{\sigma}} c 
= \bot \neq \pair{\tilde{x},\tilde{x}} = 
a \widehat{\tilde{\sigma}} (b \widehat{\tilde{\sigma}} c)$,
where $\widehat{\tilde{\sigma}}$ is the extension of 
$\tilde{\sigma}$ from Definition~\ref{def:associative}.

Even if Rabi and Sherman's proof were valid, their claim would not be
particularly useful to them, as the \weakaowfsfootnote\ they
construct~\cite[proof of Theorem~5]{rab-she:j:aowf} do not in general
have domains that are in~P\@. In contrast, our $\sigma$ {\em does\/}
have a domain that is in~P, and their method (corrected to 
remove the ``$c$'' problem) does preserve 
associativity (note: we did not say weak associativity), 
and so is useful to us.\protect\label{f:lifting}}
The fact that $\sigma$ is an AOWF (rather than merely an \weakaowf)
helps us avoid the key problem in Rabi and Sherman's extension attempt
(see Footnote~\ref{f:lifting}).

Fix any string $x_0 \not\in A$ (one must exist, since $A \not\in \p$).
Let $a_0$ be the pair $\pair{x_0 , 1x_0}$. Note that $a_0$ is neither of
the form $\pair{x,x}$ for any~$x \in \sigmastar$, nor of the form
$\pair{x,w}$ for any $x \in \sigmastar$ and any witness $w \in W_M(x)$
(because $x_0 \not\in A$ and thus does not have any witnesses). Note
that, by
the definition of~$\sigma$, 
for each~$y$, $(a_0, y) \not\in \mbox{\rm domain}(\sigma)$ and
$(y, a_0) \not\in \mbox{\rm domain}(\sigma)$.
Define the total function $\tau : \sigmastar \times
\sigmastar \rightarrow\, \sigmastar$ as follows: Whenever $(a,b) \in
\mbox{\rm domain}(\sigma)$, define $\tau(a,b) \equalsdef \sigma(a,b)$;
otherwise, define $\tau(a,b) \equalsdef a_0$. 

$\tau$ is a strong, total, commutative AOWF\@.
In particular, $\tau$ is honest, since for $a_0$, which is the
only string in the range of $\tau$ that is not in the range of
$\sigma$, it holds that $\tau(a_0 , a_0) = a_0$ and 
$|a_0| + |a_0| \leq 2|a_0|$.  Also, $\tau \in \fp$, since $\sigma \in \fp$ and
$\mbox{\rm domain}(\sigma) \in \p$. That $\tau$ is 
strong follows from the facts that 
$\mbox{\rm range}(\sigma) \seq \mbox{\rm range}(\tau)$ and $\sigma$ is 
strong. Finally, to see that
$\tau$ is associative, note that 
if $a \widehat{\sigma} (b \widehat{\sigma} c)
 = \bot$
then $a \tau (b \tau c) = a_0$ and otherwise
$a \tau (b \tau c) = a \widehat{\sigma} (b \widehat{\sigma} c)$. 
Similarly, if $(a \widehat{\sigma} b) \widehat{\sigma} c = \bot$ then 
$(a \tau b) \tau c = a_0$ and otherwise
$(a \tau b) \tau c = (a \widehat{\sigma} b) \widehat{\sigma} c$. 
The associativity of $\tau$ now follows easily, given that 
$\sigma$ is associative. The commutativity of $\tau$ is immediate
from the definition of $\tau$ and the commutativity of $\sigma$
(recall our definition of commutativity is based on (complete)
equality, and thus $(a,b) \in \mbox{\rm domain}(\sigma)$ if and only if
$(b,a) \in \mbox{\rm domain}(\sigma)$).
Hence, $\tau$ is a strong, total, commutative AOWF\@.~\qed

Rabi and Sherman emphasize the importance of explicitly exhibiting 
strong, total \weakaowfs~\cite{rab-she:j:aowf}, 
since the cryptographic protocols given
in~\cite{rab-she:j:aowf} rely on their 
existence, and they also pose as an open issue the problem of whether
a strong, total \weakaowf\ can be constructed
from any given one-way function~\cite{rab-she:t:aowf}.
The proof of Theorem~\ref{thm:aowf-equ} solves these open issues.
Indeed, the function
$\tau$ defined in the above proof shows how to construct a strong, total,
commutative
AOWF (equivalently, a strong, total, commutative \weakaowf) 
based on any clocked NP machine accepting a
language in $\np - \p$. Similarly, the proof of 
Theorem~\ref{thm:aowf-equ}
shows how, given any 
one-way function (along with its polynomial runtime and honesty bounds),
one can obtain a clocked NP machine accepting a language in $\np - \p$.
Thus, as the title of this paper claims, from any given one-way function
one can create a strong, total, commutative AOWF 
(equivalently, a strong, total, commutative \weakaowf).

Finally, we mention briefly the issue of injective (i.e., one-to-one) 
AOWFs and \weakaowfs.
Valiant's class UP (unambiguous polynomial time~\cite{val:j:checking},
see Footnote~\ref{f:lifting})
has long played a central role in complexity-theoretic 
cryptography.
Rabi and Sherman give no evidence that injective \weakaowfs\
might exist. In fact, they prove that no total \weakaowf\
can be injective.
Thus, in light of Proposition~\ref{prop:ass}.\ref{prop:ass2},
no total AOWF can be injective.
However, as Theorem~\ref{thm:injective-equ} we show that 
$\p \neq \up$ if and only if 
injective \weakaowfs\ (and indeed injective AOWFs) 
exist.

Is the lack of injectivity for total commutative AOWFs 
and \weakaowfs\
an artifact
of commutativity? Consider any commutative
function $\sigma$ such that there exist elements $a$ and $b$ with $a
\neq b$ and $(a,b) \in \mbox{\rm domain}(\sigma)$. Then $\sigma(a,b) =_c
\sigma(b,a)$, and so $\sigma$ is not injective. Now let us generalize
the notion of injectivity so as to keep the general intuition of its
behavior, yet so as to not to clash so strongly with commutativity. Given
any binary function $\sigma : \sigmastar \times \sigmastar
\rightarrow\, \sigmastar$, we say $\sigma$ 
is {\em unordered-injective\/} if and 
only if for all $a,b,c,d \in \sigmastar$, if
$(a,b), (c,d) \in \mbox{\rm domain}(\sigma)$ and $\sigma(a,b) =_c
\sigma(c,d)$, then $\{a,b\} = \{c,d\}$. That is, each element $x =_c
\sigma(a,b)$ in the range of $\sigma$ has at most one unordered pair
$\{a,b\}$ (possibly degenerate, i.e., $\{a,a\} = \{a\}$) as its
preimage. If $\sigma$ is commutative, then both orderings of this
unordered pair, $(a,b)$ and $(b,a)$, will map to~$x$; if not, one
cannot know (i.e., $\sigma(a,b) =_c x$ but $\sigma(b,a) =_c y \neq x$ is
possible).

\protect\begin{theorem}
\label{thm:injective-equ}
The following are equivalent.\footnote{\protect\singlespacing
{\bf Proof of Theorem~\ref{thm:injective-equ}.} 
That (\ref{thm:injective-equ2}) implies~(\ref{thm:injective-equ1})
follows immediately
by standard techniques,
and by Proposition~\ref{prop:ass}.\ref{prop:ass1},
(\ref{thm:injective-equ3}) implies (\ref{thm:injective-equ2}). 
That (\ref{thm:injective-equ1}), (\ref{thm:injective-equ4}), and 
(\ref{thm:injective-equ5}) are pairwise equivalent follows as a corollary
from the proof of Theorem~\ref{thm:aowf-equ} (note, crucially, that if 
the definition of $\sigma$ given in that proof
is based on some set $A \in \up - \p$, then $\sigma$ is
unordered-injective, since no string $x$ in $A$ can have more than one
witness).
So it suffices to prove that (\ref{thm:injective-equ1})
implies (\ref{thm:injective-equ3}). Assuming $A \in \up - \p$, 
define the language $A' \equalsdeffootnote \{ 1x \condition x \in
A\}$.  Clearly, $A' \in \up - \p$. Let $M$ be some UP machine
accepting~$A'$. Let the polynomial $p$ and, for each~$x$, let the 
witness sets $W_M(x)$ be defined as in the proof of Theorem~\ref{thm:aowf-equ}
(note that, for each $x \in A'$, $W_M(x)$ now is a singleton).
Without loss of generality, assume that for each $x \in A'$, 
the unique witness $w$ 
certifying that $x \in A'$ starts with a 1 as its first bit, i.e.,
$w \in 1\sigmastar$. Define the binary function $\sigma : \sigmastar \times
\sigmastar \rightarrow\, \sigmastar$ as follows:
\[
\sigma(a,b) \equalsdeffootnote \left\{ 
\begin{array}{ll} 
  0a & \mbox{if $a \in A'$ and $W_M(a) = \{b\}$} \\

  \mbox{\rm undefined} & \mbox{otherwise.}
\end{array} 
\right.
\]

Let $\widehat{\sigma}$ be the extension of $\sigma$ as in
Definition~\ref{def:associative}. Note that for all $a,b,c \in
\sigmastar$, it holds that $(a \widehat{\sigma} b) \widehat{\sigma} c
= \bot = a \widehat{\sigma} (b \widehat{\sigma} c)$ by definition
of~$\sigma$. Thus, $\sigma$ is associative according to
Definition~\ref{def:associative}. Also, $\sigma$ clearly is injective,
and the standard proof approach (see, e.g., the proof of
Theorem~\ref{thm:aowf-equ}) shows that $\sigma$ is a one-way
function.~\qed
\label{f:proof-injective}
}
{\singlespacing
\begin{enumerate}
\item \label{thm:injective-equ1} $\p \neq \up$.
\item \label{thm:injective-equ2} There exist injective 
\weakaowfsmath.
\item \label{thm:injective-equ3} There exist injective AOWFs.
\item \label{thm:injective-equ4} There exist strong, commutative,
unordered-injective \weakaowfsmath.
\item \label{thm:injective-equ5} There exist strong, commutative,
unordered-injective AOWFs.
\end{enumerate}
}
\end{theorem}

\section{Conclusions}

So, in this paper, we have shown that 
$\p \neq \np$ is a sufficient condition for 
strong, total, commutative 
AOWFs (equivalently, for strong, total, commutative 
\weakaowfs) 
to exist. Since by 
standard techniques
(namely, the natural binary-function injectivity-not-required 
analog
of a result of Grollmann and
Selman~\cite{gro-sel:j:complexity-measures,sel:j:one-way}, see
also~\cite{ko:j:operators}), 
$\p \neq \np$ is also a necessary
condition for the existence of such functions, we obtain a complete
characterization.  This characterization solves the conjecture of Rabi
and Sherman that strong \weakaowfs\
exist~\cite{rab-she:j:aowf}, inasfar as one can solve it
without solving the $\p \stackrel{\mbox{\protect\scriptsize\rm ?}}{=}
\np$ question.
Moreover, our proofs have shown how to construct
a strong, total, commutative 
AOWF (equivalently, a strong, total, commutative \weakaowf)
from any given one-way function,
which resolves an open problem of Rabi and Sherman~\cite{rab-she:t:aowf}.

We mention that most cryptographic applications are in general concerned with
average-case complexity and randomized algorithms instead of
worst-case complexity and deterministic algorithms.
However, as Rabi and Sherman stress, the intriguing concept of
(weakly) associative one-way functions,
particularly when they are total and strong
and ideally in an average-case version, may be expected to be useful in
many cryptographic applications such as in the key-agreement protocol
proposed by Rivest and Sherman in 1984 (see~\cite{rab-she:j:aowf}),
and may eventually offer elegant solutions to a variety of practical
cryptographic problems. 

\bigskip

{\samepage
\noindent 
{\bf Acknowledgments.} \quad We thank Alan Selman for 
sharing with us his knowledge of
the history and literature of partial functions,
and of Kleene's work.

}

{
\singlespacing


}

\clearpage

\end{document}